\documentstyle[12pt]{article}
\newif\ifall\alltrue
\def\private#1{}			

\def\nq{\hspace{-1em}}

\def\ignore#1{}

\def\hbar{h\!\!\!\!^{-}\,}

\def\beq{\begin{equation}}
\def\eeq{\end{equation}}
\def\bqa{\begin{eqnarray}}
\def\eqa{\end{eqnarray}}

\begin{document}
\ifall

\begin{titlepage}
\hspace*{13cm}LMU 95-..

\hspace*{13cm}May 1995

\begin{center}					  \vspace*{3cm}
{\LARGE The Mass of the $\eta'$                  }\\[0.5cm]
{\LARGE  in selfdual QCD                         }\\[3cm]
  {\bf Marcus Hutter\footnotemark}                \\[2cm]
  {\it Sektion Physik der Universit\"at M\"unchen}\\
  {\it Theoretische Physik}                       \\
  {\it Theresienstr. 37 $\quad$ 80333 M\"unchen} \\[2cm]
\end{center}
\footnotetext{E--Mail:hutter@hep.physik.uni--muenchen.de}

\begin{abstract}
The QCD gauge field is modeled as an ensemble of statistically
independent selfdual and antiselfdual regions.
This model is motivated from instanton physics.
The scale anomaly then allows to relate the topological susceptibility
to the gluon condensate. With the help of Wittens formula for
$m_{\eta'}$ and an estimate of the suppression of the gluon
condensate due to light quarks the mass of the $\eta'$ can
be related to $f_\pi$ and the physical gluon condensate.
We get the quite satisfactory value $m_{\eta'}=884\pm116$ MeV.
Using the physical $\eta'$ mass as an input it is in princple
possible to get information about the interaction between
instantons and anti-instantons.
\end{abstract}
\end{titlepage}
\stepcounter{page}


\fi
\unitlength=1.00mm
\section{Introduction}\label{sec1}

In many channels a direct calculation of the meson correlators
in the instanton liquid model and a spectral fit lead to
reasonable results for the masses of the lightest mesons \cite{ShV}.
This method even works in the axial triplet
channel because the model correctly describes spontaneous breaking
of chiral symmetry. In the axial
singlet channel a strong repulsion prevents the formation
of a meson \cite{Hut2,Geshkenbein}.
The conclusion is, that there is no massless Goldstone
boson in this channel, but the mass of the $\eta'$ remains undetermined.
In this letter I want to calculate the mass of the $\eta'$
by combining quite different techniques.
With the help of
\begin{itemize}\parskip=0ex\parsep=0ex\itemsep=0ex
\item current algebra theorems for the $\eta'$,
\item $1/N_c$ expansion,
\item instanton model,
\item scale anomaly
\end{itemize}
we are able to relate the $\eta'$ mass to the pion coupling constant
$f_\pi$ and the physical gluon condensate.

\section{Wittens formula}\label{sec2}

In leading order in $1/N_c$ it is possible to relate the $\eta'$ mass
to the $\Theta$ dependence of the topological susceptibility\footnote{
  $\langle AB\rangle_{conn} = \langle AB\rangle -
   \langle A\rangle\langle B\rangle$}
$d^2E/d\Theta^2$ of QCD without quarks \cite{Wit}:
\beq\label{pro55}
  m_{\eta'}^2 = {4N_f\over f_\pi^2}
    \left({d^2E\over d\Theta^2}\right)_{\Theta=0}^{no\;quarks}
  \quad,\quad
  {d^2E\over d\Theta^2} = \int\!d^4\!x\;
    \langle 0|{\cal T}Q(x)Q(0)|0\rangle_{conn}
\eeq
$$
  Q(x) =  {\alpha_s\over 4\pi}\mbox{tr}_c G\tilde G(x) \quad,\quad
  Q = \int\!d^4\!x\,Q(x) \in Z\!\!\!Z
$$
$Q(x)$ is the topological charge density and $Q$ the total charge.
This formula is derived by arguing, that for large
$N_c$ the topological susceptibility is dominated by the $\eta'$
state, utilizing the axial anomaly \cite{ABJ}
and the relation $f_\pi=f_{\eta'}$,
which is exact for $N_c\to\infty$.

\section{Selfdual QCD}\label{sec3}

The next step is to relate the topological susceptibility to the
gluon condensate $\langle 0|N(0)|0\rangle$:
$$
  N(x) =  {\alpha_s\over 4\pi}\mbox{tr}_c GG(x) \quad,\quad
  N = \int\!d^4\!x\,N(x)
$$
In instanton models the gluon field consists of
instantons of charge $Q=\pm 1$. The
exact N instanton solutions $(Q=N)$
are selfdual ($G_{\mu\nu}\!=\!\tilde G_{\mu\nu}$) and
\beq\label{pro80}
    \langle 0|{\cal T}Q(x)Q(0)|0\rangle_{conn} =
    \langle 0|{\cal T}N(x)N(0)|0\rangle_{conn} \quad.
\eeq
The exact anti-instanton solutions $(Q=-N)$
are anti-selfdual ($G_{\mu\nu}\!=\!-\tilde G_{\mu\nu}$) and
(\ref{pro80}) holds too, because the two minus signs cancel.
Unfortunately these exact solutions are not the most
important contributions to the partition function.

The dominating configurations are instantons and anti-instantons
in mixed combination. The simplest model is a dilute sum
$A=\sum_I A_I$ of instantons of mixed charge.
$G_{\mu\nu}$
is then approximately selfdual near the instanton centers,
approximately anti-selfdual near the anti-instanton centers and
small far away from any instanton.
In leading order in the instanton density we have
\beq
   G_{\mu\nu}(x)=\pm\tilde G_{\mu\nu}(x)
\eeq
where the sign now depends on $x$!
Let us define $N^\pm(x)$ in the following way:
$$
  N(x)=N_+(x)+N_-(x) \quad,\quad Q(x)=N_+(x)-N_-(x)
$$
$N_+(x)$ is a sum of bumps near the centers of instantons and
$N_-(x)$ near the centers of anti-instantons.
(\ref{pro80}) has to be replaced by the relation
\beq\label{pro81}
    \langle 0|{\cal T}Q(x)Q(0)|0\rangle_{conn} =
    \langle 0|{\cal T}N(x)N(0)|0\rangle_{conn} -
    4\langle 0|{\cal T}N_+(x)N_-(0)|0\rangle_{conn}
\eeq
Assuming independence of instantons and anti-instantons
$(\langle N_+N_-\rangle=\langle N_+\rangle\langle N_-\rangle)$
the equation reduces again to (\ref{pro80}). We will see that
this is a crucial assumption.

\section{The Scale Anomaly}\label{sec4}

The next ingredient is the scaling behaviour of QCD.
Classical chromodynamics is scale invariant and the Noether theorem
leads to a conserved scale current. In quantum theory the scale
invariance is anomalously broken (like the axial singlet current).
Ward identities can be derived, especially \cite{Shi}
\beq\label{pro82}
  \int\!d^4\!x\;\langle 0|{\cal T}N(x)N(0)|0\rangle_{conn} =
  {4\over b}\langle 0|N(0)|0\rangle \quad,\quad
  b = {11\over 3}N_c
\eeq
Therefore in a self(anti)dual background the topological susceptibility
$d^2E/d\Theta^2$ is proportional to the gluon condensate:
\beq\label{pro83}
  {d^2E\over d\Theta^2} = {4\over b}\langle 0|N(0)|0\rangle
\eeq
The relation is still valid, when there are statistically
independent regions
of selfduality and selfantiduality, as discussed above.
It is in fact sufficient to assure independence of the
total instanton/anti-instanton number $N_\pm$.
I have checked (\ref{pro83}) by using the theoretical one loop instanton
density $D(\rho)$ calculated in \cite{tHo}. Only the $\rho$ dependence
$D(\rho)\sim\rho^{-5}(\rho\Lambda)^b$ is important.
Due to the infrared divergence
it is neccessary to introduce an infrared cutoff, but
one has to assure not to break scale invariance.
A minimal change is to introduce two cutoffs $f_\pm$ in the total
instanton/anti-instanton packing fraction.
The packing fraction is the spacetime volume
ocupied by the instantons and is a dimensionless quantity.
Scale invariance and independence of instantons and anti-instantons
are ensured. The partition function $Z$ is
\beq\label{pro84}
  Z=\sum_{N_+N_-}Z_{N_+}^+Z_{N_-}^-
\eeq
$$
  Z_{N_\pm}^\pm = {V_4^{N_\pm}\over N_\pm!} \int_0^\infty\nq
  d\rho_1\ldots d\rho_{N_\pm}\,D(\rho_1)\ldots D(\rho_{N_\pm})\,
  \Theta\left(f_\pm-{1\over V_4}\sum_{i=1}^{N_\pm}\rho_i^4\right)
$$
A lengthy, but quite standard calculation of statistical physics,
leads to
$$
  Z_{N_\pm}^\pm = \left({c_\pm N_\pm\over V_4\Lambda^4}\right)^{
                  -{bN_\pm\over 4}}
$$
where $c_\pm=c_\pm(f_\pm,b)$ are constants independent of $N_\pm$.
Differentiation of $\ln Z$ w.r.t. $\ln c_\pm$ two times leads to
(\ref{pro83}). The result is independent of $f_\pm$.
An attractive interaction would lower the susceptibility
(compared to the density). This can be seen in the following way:
In the extrem case of a very attractive interaction, all instantons
will be
bound to intanton-anti-instanton molecules, thus $N_+=N_-$ and
$Q\equiv 0$. On the other hand a repulsive interaction
would increase the susceptibility:
\bqa\label{pro85}
  {d^2E\over d\Theta^2} &<& {4\over b}\langle 0|N(0)|0\rangle
  \quad\mbox{for attractive $I\overline{I}$ interaction} \\
  \label{pro86}
  {d^2E\over d\Theta^2} &>& {4\over b}\langle 0|N(0)|0\rangle
  \quad\mbox{for repulsive $I\overline{I}$ interaction}
\eqa
Therefore the violation of (\ref{pro83}) is a measure of
the $I\overline{I}$ interaction.

All this should be compared to Dyakonov \cite{Dya},
where the relation
\beq\label{pro57}
  d^2E/d\Theta^2=\langle 0|N(0)|0\rangle
\eeq
has been derived, which is similar to (\ref{pro83}). In this work
a simple sum ansatz $A=\sum A_I$ has been made. This ansatz
leads to a strong repulsion between close instantons,
which is the origin of the missing factor $4/b$.
The result on its own and the comparison with
(\ref{pro86}) shows that some repulsive interaction is
at work. Verbaarshot \cite{Ver2} has shown that this repulsion
is an artefact of the simple sum ansatz. Using the much more
accurate and elaborate streamline ansatz he showed, that
the interaction strongly depends on the orientation and
the average interaction is about 14 times smaller than those
obtained in \cite{Dya}.
Therefore the best thing one can do
today is to assume no intercation at all and cut the packing
fraction at some small value. There is also a more general argument that
the relation derived in \cite{Dya} must be wrong.
The topological susceptibility is of $O(N_c^0)$, whereas the
gluon condensate is of $O(N_c)$. (\ref{pro83})
is consistent with $N_c\to\infty$ considerations
only due to the presence of the $4/b$ factor.

\section{The Mass of the $\eta'$ meson}\label{sec5}

Let us now continue with the calculation of $m_{\eta'}$.
The physical gluon condensate in the presence of light quarks is
\beq
  \langle 0|N(0)|0\rangle_{phys} = (200\mbox{MeV})^4 \quad.
\eeq
Due to the presence of light quarks it is reduced by a factor $\alpha<1$.
\beq\label{pro56}
   \langle 0|N(0)|0\rangle_{phys} =
   \alpha\langle 0|N(0)|0\rangle_{\Theta=0}^{no\;quarks}
\eeq
Combining (\ref{pro55}), (\ref{pro80}), (\ref{pro82}) and (\ref{pro56}),
we get the final formula for the $\eta'$ mass is
\beq\label{pro2}
  m_{\eta'}^2 = {4N_f\over f_\pi^2}
    {12\over 11N_c}{1\over\alpha}\langle 0|N(0)|0\rangle_{phys}
\eeq
One can see that $m_{\eta'}^2\sim 1/N_c$ because $f_\pi^2$ and
the gluon condensate are proportional to $N_c$.
The largest uncertanty lies in the determination of
$\alpha$. Using again the instanton model, $\alpha$ is the
determinant of the Dirac operator with the current quark masses replaced
by effective masses:
\beq
  \alpha = \prod_{i=u,d,s}1.34m_i^{eff}\rho \approx 0.4\ldots 0.7
\eeq
We have set the effective masses to the constituent masses
\beq
  m_u^{eff}=m_d^{eff}=300\ldots 350\;\mbox{MeV} \quad,\quad
  m_s^{eff}=400\ldots 500\;\mbox{MeV}
\eeq
and $\rho$ to the value of the instanton liquid model
($\rho=600\;\mbox{MeV}^{-1}$). This estimate is consistent with the
estimate of \cite{SVZ3}.
Inserting $N_f=3$ and $f_\pi=132\;$MeV into (\ref{pro2}) we get
\beq\label{pro1}
  m_{\eta'} = 884\pm 116\mbox{MeV}
\eeq
which is in good agreement to the experimental value of 958 MeV.
This result in turn confirms the assumption, that the interaction between
selfdual and anti-selfdual regions is small. The large uncertanty
in $\alpha$ prevents more accurate statements, but (\ref{pro57})
can definitely be excluded.

Using $m_{\eta'}$ as an input
we can determine the gluon condensate in pure QCD
\beq
  {\alpha_s\over 4\pi}\langle\mbox{tr}_c GG\rangle^{no\;quarks}
  = (246\;\mbox{MeV})^4
\eeq
where we have again set $N_f$ to 3.

\section{Conclusions}\label{sec6}

We have calculated $\eta'$ successfully in a model of seldual
QCD.
The factor $4/b$ in (\ref{pro83}) is the essential term to get
the correct $N_c$
dependence of $m_{\eta'}$ and agreement with the experimental mass.
The discussion has shown, that the $\eta'$ channel can be
an experimental device for testing the independence of
selfdual and antiselfdual regions in QCD, which is an assumption
in the simplest instanton models.
It might turm out some day that the details of instanton models
are wrong but the assumption of independent self(anti)dual
regions remain valid.


\stepcounter{section}
\addcontentsline{toc}{section}{\Alph{section} References}
\parskip=0ex plus 1ex minus 1ex

\end{document}